\documentclass[12pt]{article}
\usepackage{amssymb,amsmath,epsfig}

\begin{document}

\title{\bf Modified Holographic Dark Energy in Non-flat Kaluza$-$Klein Universe with
Varying $G$}
\author{M. Sharif \thanks {msharif.math@pu.edu.pk} and Abdul Jawad\thanks
{jawadab181@yahoo.com}\\
Department of Mathematics, University of the Punjab,\\
Quaid-e-Azam Campus, Lahore-54590, Pakistan.}

\date{}

\maketitle
\begin{abstract}
The purpose of this paper is to discuss the evolution of modified
holographic dark energy with variable $G$ in non-flat Kaluza$-$Klein
universe. We consider the non-interacting and interacting scenarios
of the modified holographic dark energy with dark matter and obtain
the equation of state parameter through logarithmic approach. It
turns out that the universe remains in different dark energy eras
for both cases. Further, we study the validity of the generalized
second law of thermodynamics in this scenario. We also justify that
the statefinder parameters satisfy the limit of $\Lambda$CDM model.
\end{abstract}

\section{Introduction}

The discovery of the accelerating expansion of the universe is a
milestone for cosmology which has deep implications for the
composition of the universe, structure formation and its fate. The
expansion of the universe shows that it is not slowing down under
normal gravity but accelerating due to an unknown component termed
as dark energy (DE), having a strong negative pressure \cite{1}.
There are many pieces of evidence for the existence of this
component of the universe other than the baryonic and non-baryonic
dark matter (DM) \cite{2} but it has no clear clue about its
identity.

The most convenient explanation for this expansion is the vacuum
energy that generates sufficient force to push matter apart
described by cosmological constant \cite{2}. However, there are two
alternative ways used extensively in order to explain this behavior.
The first approach is the work on different DE models such as
quintessence \cite{3}, K-essence \cite{4}, phantom \cite{5}, quintom
\cite{6}, tachyon \cite{7}, family of Chaplygin gas \cite{8},
holographic \cite{9,10} and new agegraphic DE \cite{11}. Among all
these models, holographic DE models are widely used. They provide
the link of the DE density to the cosmic horizon \cite{12} and has
been tested through various astronomical observations \cite{13}.

The idea of holographic DE model (HDE) can be extracted from the
holographic principle, which states that \textit{the number of
degrees of freedom of a physical system should scale with its
bounding area rather than its volume} \cite{14}. Cohen et al.
\cite{15} proposed a relation of ultraviolet (UV) $\Lambda$ and
infrared (IR) cutoffs $L$ due to the limit set by forming a black
hole in quantum field theory. In their point of view, the total
energy of the system having size $L$ is bounded by the mass of black
hole of the same size. Mathematically, it can be written as
$L^3\rho_{\Lambda}\leq~LM^2_{p}$, where $\rho_{\Lambda}$ represents
the vacuum energy density and $M_{p}=(8\pi G)^{-\frac{1}{2}}$ is the
reduced Planck mass. Thus one can deduce holographic DE \cite{9}
\begin{equation*}
\rho_{\Lambda}=3m^{2}M^{2}_pL^{-2},
\end{equation*}
here constant $3m^2$ is used for convenience and $G=G(t)$.

The variation of Newton gravitational constant $G$ with cosmic time
$t$ has been considered for discussing the evolution of DE models.
This was also used for solving the longstanding problems such as the
DM problem, the controversies of Hubble parameter value and the
cosmic coincidence problem (references therein \cite{16}).
Additionally, a lot of debate is available in literature for the
choice of IR cutoff: whether it is a Hubble horizon, or particle
horizon, or future event horizon for flat FRW universe. It was
pointed out by Li \cite{10} that the future event horizon is the
appropriate choice for IR cutoff which favors the current
observations.

The modified and multidimensional theories of gravity (including
f($R$), f($G$), f($R$,$G$) \cite{17}, f($T$) \cite{18},
Brans$-$Dicke \cite{19}, Horava$-$Lifshitz \cite{20},
Kaluza$-$\\Klein (KK) \cite{21}) is the second approach in which a
phenomenon to modify the gravitational sector or increment of
dimension takes place. In case of higher dimensional theories, KK
theory has attracted many people recently to discuss the DE puzzle.
It exists in two versions: compact (fifth dimension is length like
and it should be very small) and non-compact forms (fifth dimension
is mass like) \cite{22}. In addition, on the basis of the $N+1$
dimensional mass of the Schwarzschild black hole \cite{23}, Gong and
Li \cite{24} derived the HDE in extra dimensions (called modified
holographic dark energy (MHDE)).

A marvellous work is available which investigates the
non-interacting \cite{10,25,26} and interacting \cite{27}
possibilities of HDE with DM in flat and non-flat FRW universes. The
thermodynamical interpretation of HDE model with different IR cutoff
was also investigated by many people \cite{28,29} for non-flat FRW
universe. The proposal of a statefinder $\{r,s\}$ was given by Sahni
el al. \cite{30} for characterizing and differentiating various DE
models. Alam et al. \cite{31} proved that these parameters are a
useful tool for describing the properties of DE models. Some people
\cite{32}-\cite{35} have obtained interesting results about HDE with
the help of these parameters. Also, MHDE has been considered to
discuss the evolution of the universe \cite{24,36}. Recently Sharif
et al. \cite{37,38} have explored the evolution of interacting MHDE
with Hubble horizon and event horizon as an IR cutoff in a flat KK
universe and also examined the validity of generalized second law of
thermodynamics (GSLT).

This paper is devoted to study the evolution of non-interacting and
interacting MHDE in a non-flat KK universe. The sequence of paper is
as follows: we discuss the evolution of non-interacting and
interacting MHDE in non-flat universe in the next section. In
section \textbf{3}, the generalized second law of thermodynamics is
explored. Section \textbf{4} contains the discussion of statefinder.
We summarize our results in section \textbf{5}.

\section{Modified Holographic Dark Energy and\\ Non-flat Kaluza$-$Klein Universe}

In this section, we evaluate equation of state (EoS) parameter for
the non-interacting and interacting MHDE with DM in a compact
non-flat KK universe \cite{39} whose line element is
\begin{equation}\label{1}
ds^{2}=dt^{2}-a^{2}(t)[\frac{dr^{2}}{1-kr^{2}}
+r^{2}(d\theta^{2}+\sin\theta^{2}d\phi^{2})+(1-kr^{2})d\psi^{2}],
\end{equation}
where a(t) is the cosmic scale factor responsible for the expansion
of the universe and $k=-1,0,1$ is the spatial curvature which
corresponds to open, flat and closed universe, respectively. The
Einstein field equation in $4+1$ dimensions are
\begin{equation}\label{1a}
R_{\mu\nu}-\frac{1}{2}g_{\mu\nu}R=\kappa
T_{\mu\nu},\quad(\mu,~\nu=0,1,2,3,4),
\end{equation}
where $R_{\mu\nu},~g_{\mu\nu},~R,~T_{\mu\nu}$ and $\kappa$ represent
the Ricci tensor, the metric tensor, the Ricci scalar, the
energy-momentum tensor and the coupling constant, respectively.
Also, we assume that the KK universe is filled with perfect fluid
whose energy-momentum tensor is given by
\begin{equation}\label{1b}
T_{\mu\nu}=(p+\rho)u_{\mu}u_{\nu}-g_{\mu\nu}p,
\end{equation}
where $p=p_{\Lambda}$ is the pressure due to DE,
$\rho=\rho_{\Lambda}+\rho_m$ is the DE plus DM (dust like) energy
densities and $u_{\mu}$ is the five velocity satisfying the relation
$u^{\mu}u_{\mu}=1$, respectively. Using Eqs.(\ref{1a}) and
(\ref{1b}), we get the following Einstein field equations for
non-flat KK universe:
\begin{eqnarray}\label{2}
H^2+\frac{k}{a^{2}}&=&\frac{8\pi G}{6}\rho,\\\label{3}
\dot{H}+2H^2+\frac{k}{a^{2}}&=&-\frac{8\pi G}{3}p.
\end{eqnarray}
Here $H$ is the Hubble parameter and dot is the differentiation with
respect to time. Equation (\ref{2}) can be written in terms of
fractional energy densities as
\begin{equation}\label{4}
\Omega_{m}+\Omega_{\Lambda}=1+\Omega_{k},
\end{equation}
where
\begin{eqnarray}\label{5}
\Omega_{m}=\frac{8\pi G\rho_{m}}{6H^{2}},\quad
\Omega_{\Lambda}=\frac{8\pi G\rho_{\Lambda}}{6H^{2}},\quad
\Omega_{k}=\frac{k}{a^2H^{2}}.
\end{eqnarray}

The MHDE in $N$ dimensions is given as \cite{24}
\begin{equation*}
\rho_{\Lambda}=\frac{m^{2}(N-1)A_{N-1}L^{N-5}M^{2}_{p}}{2V_{N-3}},
\end{equation*}
where $A_{N-1}$ and $V_{N-3}$ indicate the area of unit $N$-sphere
and volume of the extra-dimensional space, respectively. In case of
the KK cosmology, it becomes
\begin{equation}\label{6}
\rho_{\Lambda}=\frac{3m^{2}\pi^{2}L^{2}}{8\pi G}.
\end{equation}
Here $L$ is defined for non-flat KK universe as \cite{25}
\begin{equation}\label{7}
L=a(t)r(t),
\end{equation}
where $r(t)$ is obtained through the relation
\begin{equation*}
\int^{r(t)}_{0}\frac{1}{\sqrt{1-kr^2(t)}}=\frac{R_e}{a(t)}\equiv u.
\end{equation*}
Further, its integration gives
\begin{eqnarray*}
r(t)=\frac{1}{\sqrt{|k|}}sinn\left(\sqrt{|k|}u\right)=\left\{\begin{array}{ll}
\sin{u}, & k=+1,\\
u, &  k=0,\\
\sinh{u}, & k=-1.\\
\end{array}\right.
\end{eqnarray*}
$R_e$ is the the future event horizon which is defined as
\begin{equation*}
R_e=a(t)\int^{\infty}_{a}{\frac{d\tilde{a}}{H\tilde{a}^{2}}}
=a(t)\int^{\infty}_{t}{\frac{d\tilde{t}}{a(\tilde{t})}}.
\end{equation*}
The time derivative of $L$ yields
\begin{equation}\label{8}
\dot{L}=HL-cosn\left(\sqrt{|k|}u\right)
\end{equation}
with
\begin{eqnarray*}
\frac{1}{\sqrt{|k|}}cosn(\sqrt{|k|}u)=\left\{\begin{array}{ll}
\cos{u}, & k=+1,\\
1, &  k=0,\\
\cosh{u}, & k=-1.\\
\end{array}\right.
\end{eqnarray*}
Using Eqs.(\ref{6}) and (\ref{8}), we obtain the evolution of energy
density
\begin{eqnarray}\label{9}
\rho'_{\Lambda}=\rho_{\Lambda}\left[2-\frac{2m\pi}{H^{2}\sqrt{2{\Omega}_{\Lambda}}}
cosn(\sqrt{|k|}\delta)-\Delta_{G}\right],
\end{eqnarray}
where $\Delta_{G}\equiv\frac{G'}{G},~\dot{G}=G'H$ and prime
represents derivative with respect to $\ln a(t)$. By differentiating
fractional energy densities $\Omega_{\Lambda}$, $\Omega_{k}$ and
using Eq.(\ref{8}), it follows that
\begin{equation}\label{10}
\Omega'_{\Lambda}=2{\Omega}_{\Lambda}(1-\frac{m\pi}
{H^{2}\sqrt{2{\Omega}_{\Lambda}}}cosn(\sqrt{|k|}\delta)-\frac{\dot{H}}{H^2}),\quad
\Omega'_{k}=-2{\Omega}_{\Lambda}(1+\frac{\dot{H}}{H^2}).
\end{equation}
In the following, we investigate EoS parameter for non-interacting
and interacting cases.

\subsection{Non-Interacting Case}

The equation of continuity for the KK universe is
\begin{equation*}
\dot{\rho}+4H(\rho+p)=0.
\end{equation*}
We consider the universe filled with DE and DM which splits it into
two equations for DM and DE, respectively, as
\begin{eqnarray}\label{11}
\dot{\rho}_{m}+4H\rho_{m}=0,\\\label{12}
\dot{\rho}_{\Lambda}+4H(\rho_{\Lambda}+p_{\Lambda})=0.
\end{eqnarray}
In order to eliminate the term including $\dot{H}$ in Eq.(\ref{10}),
we use Eqs.(\ref{2}), (\ref{6}) and (\ref{11}):
\begin{eqnarray}\label{13}
\frac{2\dot{H}}{H^2}&=&\left[-4-2{\Omega}_{k}+{\Omega}_{\Lambda}(6-\frac{2m\pi}
{H^{2}\sqrt{2\Omega_{\Lambda}}}cosn(\sqrt{|k|}\delta))+\Delta_{G}(1+\Omega_{k}\right.
-\left.\Omega_{\Lambda})\right].\nonumber\\
\end{eqnarray}
Inserting this value in Eq.(\ref{10}), it follows that
\begin{eqnarray}\nonumber
{\Omega}'_{\Lambda}&=&{\Omega}_{\Lambda}\left[(1-{\Omega}_{\Lambda})(6-\frac{2m\pi}
{H^{2}\sqrt{2\Omega_{\Lambda}}}cosn(\sqrt{|k|}\delta))
+2{\Omega}_{k}-\Delta_{G}(1+\Omega_{k}\right.\\\nonumber
&-&\left.\Omega_{\Lambda})\right],\\\nonumber
{\Omega}'_{k}&=&{\Omega}_{k}\left[2-{\Omega}_{\Lambda}(6-\frac{2m\pi}
{H^{2}\sqrt{2\Omega_{\Lambda}}}cosn(\sqrt{|k|}\delta))
+2{\Omega}_{k}-\Delta_{G}(1+\Omega_{k}\right.\\\nonumber
&-&\left.\Omega_{\Lambda})\right].\\\label{14}
\end{eqnarray}

Now we want to extract EoS parameter $w_{\Lambda}$ in terms of
redshift parameter $z$. Integration of the conservation equation for
DE gives
\begin{equation*}
\rho_{\Lambda}\sim~a^{-4(1+\omega_{\Lambda})}.
\end{equation*}
We follow the procedure of Li \cite{10} and use Taylor expansion of
the DE density around the present time $a_{0}=1$ as
\begin{equation*}
\ln\rho_{\Lambda}=\ln\rho^{0}_{\Lambda}+\frac{d\ln\rho_{\Lambda}}{d\ln
a}\ln a+\frac{1}{2}\frac{d^2\ln\rho_{\Lambda}}{d(\ln a)^2}(\ln
a)^2+...,
\end{equation*}
where $\rho^{0}_{\Lambda}$ is the present value of the DE density.
The EoS parameter, up to second order expansion, becomes
\begin{eqnarray}
\omega_{\Lambda}=-1-\frac{1}{4}\left(\frac{d\ln\rho_{\Lambda}}{d\ln
a}\right)+\frac{1}{8}\left(\frac{d^2\ln\rho_{\Lambda}}{d(\ln
a)^2}\right)z\equiv\omega_{0}+\omega_{1}z,\label{15}
\end{eqnarray}
here we have assumed the small redshift approximation, i.e.,
$\ln{a}=-\ln(1+z)\simeq~-z$, where
\begin{eqnarray}\label{16}
\omega_{0}=-1-\frac{1}{4}\left(\frac{d\ln\rho_{\Lambda}}{d\ln{a}}\right),\quad
\omega_{1}=\frac{1}{8}\frac{d^2\ln\rho_{\Lambda}}{d(\ln{a})^2}.
\end{eqnarray}
From (\ref{5}) and (\ref{11}), one can get
\begin{equation}\label{17}
\rho_{\Lambda}=\frac{\rho_{m_{0}}a^{-4}\Omega_{\Lambda}}{1+\Omega_{k}-\Omega_{\Lambda}}.
\end{equation}
Making use of Eqs.(\ref{14}), (\ref{16}) and (\ref{17}), it follows
that
\begin{eqnarray}\label{18}
\omega_{0}&=&-\frac{3}{2}+\frac{m\pi}
{2H_{0}^{2}\sqrt{2\Omega^{0}_{\Lambda}}}\sqrt{1-\frac{2H_{0}^4\Omega^{0}_{\Lambda}|\Omega^{0}_{k}|}{m^2\pi^2}}
+\frac{1}{4}\Delta_{G},\\\nonumber
\omega_{1}&=&\frac{1}{4}\left[\Omega^{0}_{\Lambda}|\Omega^{0}_{k}|
-(1+\Omega^{0}_{\Lambda})\left(\frac{m^2\pi^2}{H_{0}^4\Omega^{0}_{\Lambda}}\right)
+\frac{m\pi}{H_{0}^2\sqrt{2\Omega^{0}_{\Lambda}}}(-4-\Omega^{0}_{k}\right.\\\label{29}
&+&\left.3(1+\Omega^{0}_{\Lambda})+\frac{1}{2}\Delta_{G}(1+\Omega^{0}_{k}-\Omega^{0}_{\Lambda}))
\sqrt{1-\frac{2H_{0}^4\Omega^{0}_{\Lambda}|\Omega^{0}_{k}|}{m^2\pi^2}}\right],
\end{eqnarray}
\begin{figure} \centering
\epsfig{file=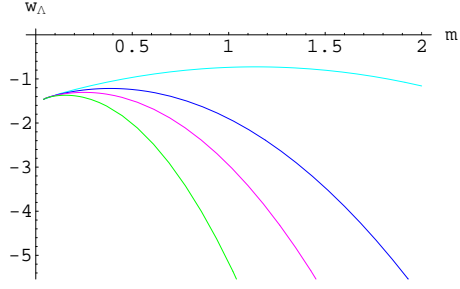,width=.45\linewidth} \caption{Plot of
$\omega_{\Lambda}$ versus $m$ for non-interacting case.}
\end{figure}
where ($0$) denotes the present time value of the parameters.

Finally, we obtain $\omega_{\Lambda}$ by inserting the above
equations in (\ref{15}). In the evolution of $\omega_{\Lambda}$, the
MHDE parameter $m$ plays a crucial role and hence we plot
$\omega_\Lambda$ against $m$ as shown in Figure \textbf{(1)}. Here
we consider the present values of fractional energy densities
$\Omega^{0}_{\Lambda}=0.73$, $\Omega^{0}_{k}=0.01$, $H_{0}=1$
\cite{40} and $0\leq\Delta_{G}\leq0.07$ \cite{16}. Plots in
turquoise, blue, pink and green curves correspond to values of
redshift parameter $z=0.1,0.31,0.5,0.9,$ respectively. We observe
that for $z=0.1$, the EoS parameter attains different phases of the
universe such as phantom ($0.04<~m<0.45$), vacuum DE
($m=0.45,1.56$), quintessence ($0.45<~m<1.56$) and then phantom for
$m>1.56$. However, for $z=0.1,0.31,0.5,0.9$, the EoS achieves its
maximum value $-0.6,-1.2,-1.25,-1.3$, respectively, and remains in
the phantom region except $z=0.1$.

\subsection{Interacting Case}

Here we evaluate EoS parameter in the interacting phenomenon of MHDE
with DM by using the same procedure as above. In this case, the
equation of continuity may be converted into two non-conserving
equations for DM and MHDE, respectively, i.e.,
\begin{eqnarray}\label{20}
\dot{\rho}_{m}+4H\rho_{m}&=&\Upsilon,\\\label{21}
\dot{\rho}_{\Lambda}+4H(\rho_{\Lambda}+p_{\Lambda})&=&-\Upsilon,
\end{eqnarray}
where $\Upsilon$ denotes the interaction term which can be taken as
\cite{27}
\begin{equation*}
\Upsilon=4d^{2}H\rho_{\Lambda}.
\end{equation*}
The parameter $d$ is a coupling constant and the selection of its
square leads to the condition of decay from DE to DM. We can find
$\omega_{0}$ and $\omega_{1}$ by using the above procedure as
follows:
\begin{eqnarray}\label{22}
\omega_{0}&=&-\frac{3}{2}-d^{2}+\frac{m\pi}
{2H_{0}^{2}\sqrt{2\Omega^{0}_{\Lambda}}}
\sqrt{1-\frac{2H_{0}^4\Omega^{0}_{\Lambda}|\Omega^{0}_{k}|}{m^2\pi^2}}
+\frac{1}{4}\Delta_{G},\\\nonumber
\omega_{1}&=&\frac{1}{4}\left[\Omega^{0}_{\Lambda}|\Omega^{0}_{k}|
-(1+\Omega^{0}_{\Lambda})\left(\frac{m^2\pi^2}{H_{0}^4\Omega^{0}_{\Lambda}}\right)
+\frac{m\pi}{H_{0}^2\sqrt{2\Omega^{0}_{\Lambda}}}(-4-\Omega^{0}_{k}+3\right.\\\label{23}
&+&\left.3\Omega^{0}_{\Lambda})+\frac{1}{2}\Delta_{G}(1+\Omega^{0}_{k}-\Omega^{0}_{\Lambda}
+2d^{2}\Omega^{0}_{\Lambda})\sqrt{1
-\frac{2H_{0}^4\Omega^{0}_{\Lambda}|\Omega^{0}_{k}|}{m^2\pi^2}}\right].
\end{eqnarray}
\begin{figure} \centering
\epsfig{file=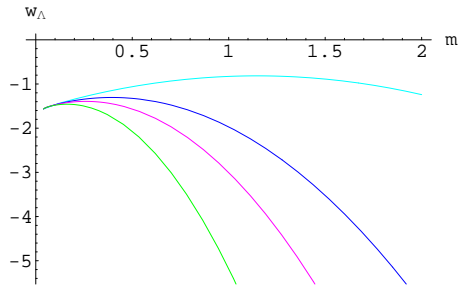,width=.45\linewidth} \caption{Plot of
$\omega_{\Lambda}$ versus $m$ for interacting case.}
\end{figure}
Inserting these values in Eq.(\ref{15}), we obtain EoS parameter. In
this case, $\omega_{\Lambda}$ evolves two constant parameters $m$
and $d$ (interacting parameter). Figure \textbf{(2)} shows the plot
of $\omega_{\Lambda}$ against $m$ by setting $d^{2}=0.1$ and the
remaining parameters are the same as in the non-interacting case.
Notice that the maximum values of EoS parameter corresponding to the
redshift parameter becomes smaller than that of the non-interacting
case.
\section{Generalized Second Law of Thermodynamics}

Now we investigate the validity of GSLT for non-interacting and
interacting MHDE with varying G in the non-flat KK universe.
According to this law, the sum of entropy of matter inside and at
the event horizon remains always positive with the passage of time
\cite{41}. Thermodynamics of black hole plays the role of pillar for
thermodynamical interpretation of the universe. Bekenstein \cite{42}
suggested that, in view of the proportionality relation between
entropy of black hole horizon and horizon area, the sum of black
hole entropy and the background entropy must be an increasing
quantity with time. The first law of thermodynamics gives
\begin{equation}\label{24}
TdS=pdV+dE,
\end{equation}
where $T,~S,~E$ and $p$ represent temperature, entropy, internal
energy and pressure of the system, respectively. Splitting this law
for DE, DM and differentiating with respect to time, we obtain
\begin{equation}\label{25}
\dot{S_{\Lambda}}=\frac{p_{\Lambda}\dot{V}+\dot{E_{\Lambda}}}{T},\quad
\dot{S_{m}}=\frac{p_{m}\dot{V}+\dot{E_{m}}}{T}.
\end{equation}
The volume, temperature and entropy of horizon in KK universe become
\cite{43}
\begin{equation}\label{26}
V=\frac{\pi^{2} L^{4}}{2},\quad T=\frac{1}{2\pi L},\quad
S_{H}=\frac{\pi^{2}L^{3}}{2G}.
\end{equation}
Also, we require the following thermodynamical quantities:
\begin{eqnarray}\label{27}
E_{\Lambda}=\frac{\pi^{2} L^{4}\rho_{\Lambda}}{2},\quad
E_{m}=\frac{\pi^{2} L^{4}\rho_{m}}{2}.
\end{eqnarray}
Equation (\ref{4}) can be re-written as
\begin{equation}\label{28}
\frac{\rho_{m}}{\rho_{\Lambda}}=-1+\frac{1}{\Omega_{\Lambda}}+\frac{\Omega_k}{\Omega_\Lambda}.
\end{equation}

In view of the above equations, we have
\begin{eqnarray}\nonumber
\dot{S}_{total}&=&\frac{3m^2\pi^4L^6}{2G}
\left[\left(-\omega_{\Lambda}-\frac{1}{\Omega_{\Lambda}}-\frac{\Omega_k}{\Omega_\Lambda}
-\frac{m^2\pi^2}{4H^{4}\Omega^{2}_{\Lambda}}\right)\sqrt{1
-\frac{2H^4\Omega_{\Lambda}|\Omega_{k}|}{m^2\pi^2}}\right.\\\label{29}
&+&\left.\frac{m\pi}{3(2\Omega_{\Lambda})^{\frac{3}{2}}H^2}(3-\Delta_{G})\right],
\end{eqnarray}
where $S_{total}$ is the sum of three entropies. When we substitute
the present values of $\Omega_{\Lambda},~\Omega_{k}$,
$\Delta_{G},~\omega_{\Lambda}$ (with $z\leq0.75$) and $m\geq0.1$ in
the above expression, it remains non-negative for both interacting
and non-interacting cases of MHDE i.e., $\dot{S}_{total}\geq0$.

\section{The Statefinder Diagnostic}

Here we explore the behavior of statefinder parameters in the above
mentioned scenario. These parameters have geometrical diagnostic due
to their total dependence on expansion factor. These are defined for
a non-flat KK universe as \cite{44}
\begin{eqnarray}\label{30}
r=\frac{\dddot{a}}{aH^3},\quad
s=\frac{r-\Omega_{tot}}{3(q-\frac{\Omega_{tot}}{2})},
\end{eqnarray}
where $\Omega_{tot}=\Omega_{\Lambda}+\Omega_{m}=1+\Omega_{k}$ and
$q$ is the deceleration parameter defined as
\begin{eqnarray}\label{31}
q=-\frac{\ddot{a}}{aH^2}.
\end{eqnarray}
The statefinder parameters are dimensionless and exhibit expansion
of the universe through higher derivatives of the scale factor.
These are a natural companion to the deceleration and Hubble
parameters. The pair $(r,s)$ defines the well-known $\Lambda$CDM
model at the fixed point $(r,s)=(\Omega_{tot},0)$. Moreover, $r$ can
be expressed in terms of the Hubble parameter as
\begin{equation}\label{32}
r=\frac{\ddot{H}}{H^3}-3q-2.
\end{equation}
With the help of Eqs.(\ref{31}) and (\ref{32}), one can write
\begin{equation}\label{33}
r=2q^{2}+q-\frac{\dot{q}}{H}.
\end{equation}

In the non-interacting case, the time derivative of the deceleration
parameter becomes
\begin{equation*}
\dot{q}=H[q(-2+\Delta_G+2q)-4\Omega_{\Lambda}(1+2\omega_{\Lambda})+2\Omega_{\Lambda}\omega'_{\Lambda}].
\end{equation*}
Inserting this in Eq.(\ref{33}), we obtain
\begin{eqnarray}\nonumber
r&=&(3-\Delta_{G})[(1+\Omega_k)(1-\frac{1}{2}\Delta_{G})+2\omega_{\Lambda}\Omega_{\Lambda}]
+4\omega_{\Lambda}\Omega_{\Lambda}\\\label{34}
&\times&(1+2\omega_{\Lambda}-\frac{1}{2}\Delta_{G})-2\Omega_{\Lambda}\omega'_{\Lambda},\\\nonumber
s&=&\frac{(3-\Delta_{G})[(1+\Omega_k)(1-\frac{1}{2}\Delta_{G})+2\omega_{\Lambda}\Omega_{\Lambda}]
+4\omega_{\Lambda}\Omega_{\Lambda}}
{3[(1+\Omega_k)(1-\frac{1}{2}\Delta_{G})+2\omega_{\Lambda}\Omega_{\Lambda}-\frac{\Omega_{tot}}{2}]}\\\label{35}
&+&\frac{4\omega_{\Lambda}\Omega_{\Lambda}(2\omega_{\Lambda}
-\frac{1}{2}\Delta_{G})-2\Omega_{\Lambda}\omega'_{\Lambda}-\Omega_{tot}}
{3[(1+\Omega_k)(1-\frac{1}{2}\Delta_{G})+2\omega_{\Lambda}\Omega_{\Lambda}-\frac{\Omega_{tot}}{2}]},
\end{eqnarray}
For the interacting case, we differentiate Eq.(\ref{3}), using
(\ref{9}) and (\ref{13}), and it follows that
\begin{eqnarray*}\nonumber
\frac{\ddot{H}}{H^3}&=&8+5\Omega_{k}+8\omega_{\Lambda}\Omega_{\Lambda}-2\Delta_{G}(1+\Omega_{k}
+\Omega_{\Lambda}\omega_{\Lambda})\\
&+&8\Omega_{\Lambda}\omega_{\Lambda}(1+\omega_{\Lambda})
-2\Omega_{\Lambda}\omega'_{\Lambda}+8d^2\Omega_{\Lambda}.
\end{eqnarray*}
\begin{figure} \centering
\epsfig{file=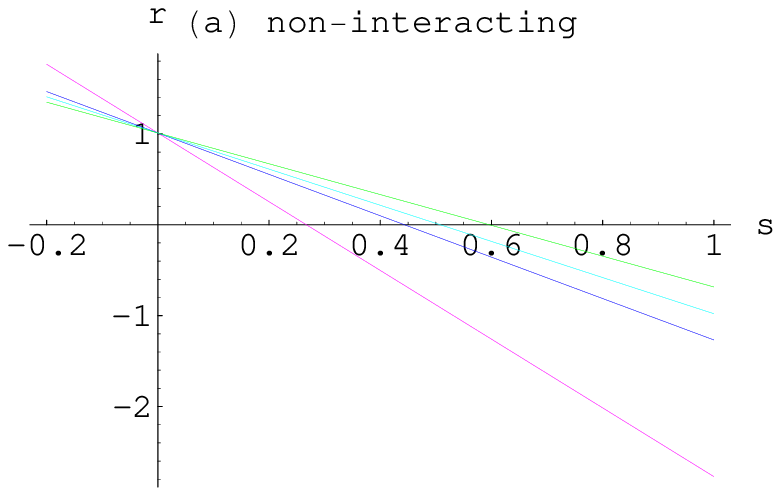,width=.45\linewidth}
\epsfig{file=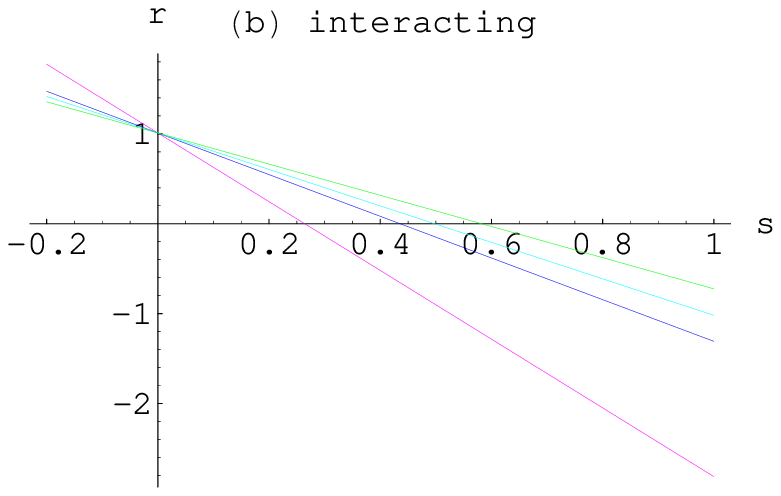,width=.45\linewidth} \caption{Plots of $r$ versus
$s$ for (a) non-interacting and (b) interacting cases.}
\end{figure}
Consequently, the corresponding statefinder takes the form
\begin{eqnarray}\nonumber
r&=&(3-\Delta_{G})[(1+\Omega_k)(1-\frac{1}{2}\Delta_{G})+2\omega_{\Lambda}\Omega_{\Lambda}]
+4\omega_{\Lambda}\Omega_{\Lambda}\\\label{36}
&\times&(1+2\omega_{\Lambda}-\frac{1}{2}\Delta_{G})-2\Omega_{\Lambda}\omega'_{\Lambda}
+8d^2\Omega_{\Lambda},\\\nonumber
s&=&\frac{(3-\Delta_{G})[(1+\Omega_k)(1-\frac{1}{2}\Delta_{G})+2\omega_{\Lambda}\Omega_{\Lambda}]
+4\omega_{\Lambda}\Omega_{\Lambda}}
{3[(1+\Omega_k)(1-\frac{1}{2}\Delta_{G})+2\omega_{\Lambda}\Omega_{\Lambda}-\frac{\Omega_{tot}}{2}]}\\\label{37}
&+&\frac{(2\omega_{\Lambda}-\frac{1}{2}\Delta_{G})
-2\Omega_{\Lambda}\omega'_{\Lambda}+8d^2\Omega_{\Lambda}-\Omega_{tot}}{3[(1+\Omega_k)(1-\frac{1}{2}\Delta_{G})
+2\omega_{\Lambda}\Omega_{\Lambda}-\frac{\Omega_{tot}}{2}]}.
\end{eqnarray}
We can easily find a single relation of $r$ in terms of $s$ and draw
$s-r$ plane as shown in Figure \textbf{(3)} for ($a$)
non-interacting and ($b$) interacting MHDE. Plots in pink, blue,
turquoise and green colors are drawn at different physically
acceptable values of $m=0.21,0.61,0.73,0.91$ (as already discussed
in \cite{38}), respectively, for non-interacting as well as
interacting cases. Also, we fix $z=0.1,d^2=0.1$ and recover the
$\Lambda$CDM model in both cases.

\section{Concluding Remarks}

We have investigated the behavior of EoS parameter, GSLT and
statefinder for the MHDE (non-interacting and interacting with DM)
with variable $G$ correction in non-flat KK universe enclosed by
future event horizon. Actually, the MHDE exhibits the dynamical
nature of the vacuum DE through its parameter $m$. We have evaluated
the EoS parameter with respect to $m$ for different ranges of
$z=0.1,0.31,0.5,0.9$. It is found that the interacting and
non-interacting MHDE behave like a quintom model for a comparatively
smaller value of the redshift parameter, i.e., $z=0.1$ with the
assumptions of the present values of the other parameters. For other
values of $z$, it evolutes the universe in phantom DE era in view of
increasing $m$. Our results about evolution of MHDE with varying $G$
shows compatibility with the present observations for flat, non-flat
FRW \cite{16,26} and flat KK universes \cite{38} enclosed by future
event horizon that it can cross the phantom divide.

Secondly, we have explored that GSLT is satisfied with $z\geq0.75$
for the universe describing phantom evolution. Moreover, we have
obtained the evolution of non-interacting and interacting MHDE in
the statefinder plane for different best fitted values of $m,~d$
(for interacting case) and $z=0.1$. The trajectories of $s$-$r$
plane have been achieved with respect to different model parameters
which is started from right to left. Notice that the parameters $s$
and $r$ show decreasing and increasing behavior, respectively, with
the phantom evolution of the KK universe.

\end{document}